\def\om{{\Omega_p}}
\def\etal{{\it et al.}}

\documentstyle[11pt,newpasp,twoside,psfig,epsf]{article}

\begin{document}

\title{Fabry-Perot Absorption-Line Spectroscopy of NGC 7079}
\author{Victor P. Debattista}
\affil{Astronomisches Institut der Universit\"at Basel, Venusstrasse 7, 
CH-4102 Binningen, Switzerland}

\author{T.B. Williams}
\affil{Dept. of Physics and Astronomy, Rutgers University, PO Box 849, 
Piscataway, NJ  08855, USA}

\begin{abstract}
We report on absorption-line Fabry-Perot spectroscopy of the SB0 galaxy NGC 
7079.  Our data allow us to determine velocity and dispersion maps, which 
compare well with the results of slit observations.  Our goals are to obtain
a bar pattern speed and to determine the velocity ellipsoid as a function 
of radius; this is work in progress.
\end{abstract}

\vspace*{-0.5cm}
\section{Introduction}
Bar pattern speeds, $\om$, constrain the dark matter content of barred galaxies
(Debattista \& Sellwood 1997).  The measurement of $\om$ in a statistically 
significant number of galaxies is, therefore, very desireable.  Tremaine \&
Weinberg (1984) derived a model-independent equation for $\om$ of systems 
satisfying the continuity equation.  Using this method with slit spectra, 
Merrifield \& Kuijken (1995) and Gerssen \etal \ (1998) found fast bars (i.e. 
corotation radius/bar semi-major axis $\simeq 1.2 \pm 0.2$) in NGC 936 and 
NGC 4596 respectively.

\vspace*{-0.25cm}
\section{Observations}
NGC 7079 was selected for observation because it has a suitably placed disk 
and bar, has several bright stars within $80''$ of the galaxy center (for 
normalizations), lacks spirals, dust and large companions, is large and 
bright, has a known recession velocity and the CaII 8542.14 \AA \ line is 
redshifted to wavelengths outside the bright sky-emission forests.  This 
galaxy, classified as (L)SB(r)0$^0$, is the brightest member of a group of 
seven galaxies (Garcia 1993).  The slit spectra of Bettoni \& Galletta (1996) 
through the center of NGC 7079 showed a small amount of counter-rotating gas 
in the inner $15''$.

We used the CTIO 0.9m telescope to obtain multiple $U$, $B$, $V$, $R$ and $I$ 
exposures of 300 $s$ each, with seeing $\sim 1.5''$.  After processing in 
the standard way, the exposures in each filter were combined.  Ellipse fits at 
large radii ($R \geq 51.5''$) gave a disk inclination, $i$, of 
$50.3^{+0.3}_{-0.4}$, which compares well with the results of Bettoni \& 
Galletta ($i = 51^\circ$).  The $B$-$I$ map shows that the disk and the bulge 
each have a constant color, indicating uniform (or zero) internal 
extinction.  Thus the Tremaine-Weinberg method can be used for this galaxy.

\begin{figure}[!t]
\centerline{\psfig{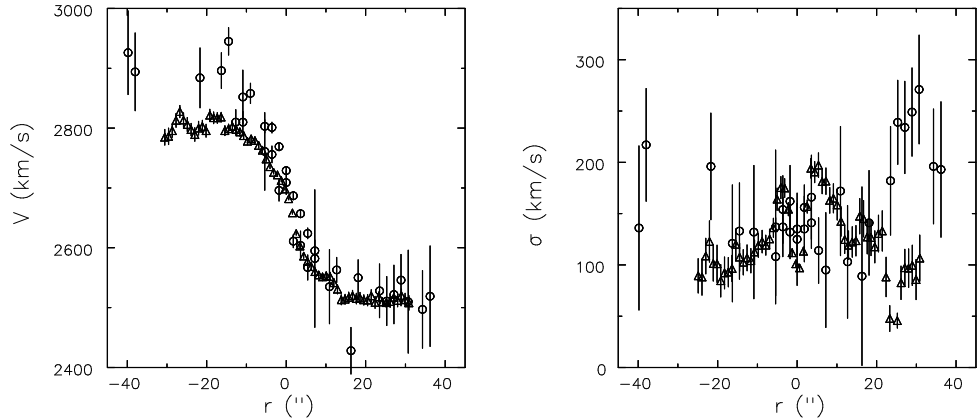}}
\caption{Our data (triangles) and those of Bettoni \& Galletta (circles) for
a slit at PA of $82^\circ$.  The effective slit width is roughly $0.7''$.}
\label{fig-1}
\end{figure}

We observed NGC 7079 with the Rutgers Fabry-Perot imaging interferometer 
on the CTIO 4m telescope, with variable seeing of $\sim 1.4'' - 2.4''$.  We 
used the CaII 8542.14 \AA \ absorption line, redshifted to 8618 \AA, scanning 
the spectrum from $8608$ \AA \ to $8631$ \AA, in steps of $1$ \AA, for a total 
of 25 exposures of 900 seconds each.  After flattening, zero subtraction and 
cosmic ray removal in the usual way, sky and sky-emission rings were 
subtracted by radial sampling in the half of the frame which excluded the 
galaxy.  The frames were normalized using bright field stars and then Voigt 
profiles were fitted to the spectrum at each pixel, giving maps of the 
velocity and dispersion.  In Figure ~\ref{fig-1} we compare our velocity and 
dispersion data with those of Bettoni \& Galletta along their major-axis slit; 
our data compare well inside $r = 20''$ from the galaxy center.  Figure 
~\ref{fig-1} shows peaks in the velocity dispersion away from the galaxy 
center.  We are still trying to understand their cause; perhaps they are 
related to the infalling material found by Bettoni \& Galletta.

\vspace*{-0.5cm}
\section{Discussion}
With the velocity and photometric data, we can now calculate $\om$ for NGC 
7079.  Since we have 2D data, we will take full advantage of the freedom in the
Tremaine-Weinberg method to reduce the uncertainty in the measurement of 
$\om$.

\end{document}